\documentclass[aps,prl,reprint,groupedaddress,superscriptaddress,floatfix]{revtex4-2}
\usepackage{amsmath}
\usepackage{amssymb}
\usepackage{braket}
\usepackage{graphicx}
\usepackage[colorlinks=true,linkcolor=blue,citecolor=blue,urlcolor=blue]{hyperref}

\begin{document}

\title{Attainable and unattainable gauge-dependent quantum precision limits in ultrastrong light--matter coupling}

\author{Wangjun Lu}
\email{wjlu1227@zju.edu.cn}
\affiliation{School of Information Science and Engineering, Hunan Institute of Engineering, Xiangtan 411104, China}

\author{Qing Yu}
\email{yq@hnie.edu.cn}
\affiliation{School of Information Science and Engineering, Hunan Institute of Engineering, Xiangtan 411104, China}

\begin{abstract}
In this paper, we study the gauge dependence of quantum precision in ultrastrongly coupled light--matter systems. The unitary between gauges carries the coupling $g$, so each truncation defines a different family. The quantum Fisher information traces an exact parabola across frames. Harmonic matter minimizes it at a universal $\theta_\ast=\Omega/(\Omega+\omega)$, set by the matter and cavity frequencies. Only the minimal-coupling bound is attainable. Frames differ by a factor $9.6$ with opposite trends. Diverging virtual-excitation gain at the Dicke point is a gauge artifact and saturates.
\end{abstract}

\maketitle

\smallskip\noindent\textit{Introduction.} Normalized light--matter couplings $g/\omega$ of order unity and beyond are now routine in circuit quantum electrodynamics~\cite{FornDiaz2019,Kockum2019}. Here $g$ is the coupling strength and $\omega$ the bare cavity frequency. In this ultrastrong-coupling (USC) regime the rotating-wave approximation fails. Counter-rotating terms and the diamagnetic $A^2$ term become essential. Even the two-level approximation becomes gauge dependent~\cite{DeBernardis2018,DiStefano2019,StokesNazir2019,StokesNazir2024}. The resulting gauge ambiguity has so far been a problem of model building. The question was which effective Hamiltonian reproduces the spectrum and matrix elements of the full light--matter Hamiltonian.

Here we show that the same ambiguity enters a quantity normally regarded as physical: the quantum Fisher information (QFI)~\cite{Braunstein1994,Paris2009,Degen2017}. The QFI sets the quantum Cram\'er--Rao (CR) bound on estimating the coupling $g$. Quantum estimation has been applied to USC models before. It was used to estimate the diamagnetic term~\cite{Rossi2017}, and to argue that virtual excitations fund precision gains beyond the standard quantum limit~\cite{Hotter2025}. In both cases the QFI was evaluated within a single gauge. We show that this choice is itself an error of order unity.

The origin of the effect is elementary but has been overlooked. The unitary connecting the two gauges is $\hat{U}=\exp(i g\,\hat{x}\hat{X})$. Here $\hat{x}$ is the matter coordinate, $\hat{a}$ the photon annihilation operator, and $\hat{X}=\hat{a}+\hat{a}^{\dagger}$ the field quadrature. The exponent carries the very parameter to be estimated. Changing gauge therefore does not merely relabel states. It defines a different family of states $\hat{\rho}(g)$. The QFI measures distinguishability within a family, so it is not invariant under such a change. The error is exactly computable, and it reverses conclusions. In the deep-strong-coupling regime the two gauges disagree on whether precision improves or degrades with coupling. We also solve the harmonic model in closed form. We then re-examine the recent critical-metrology proposal of Ref.~\cite{Hotter2025}.

\begin{figure}[!b]
\includegraphics[width=\linewidth]{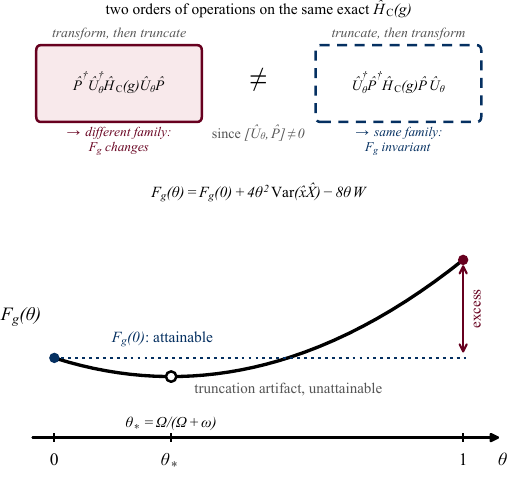}
\caption{(Color online) Top: two orders of operations on the same exact
model $\hat{H}_{\rm C}(g)$. Left: transform, then truncate. Right:
truncate, then transform; within this same family the covariant QFI
equals $F_g(0)$. The endpoints differ because
$[\hat{U}_\theta,\hat{P}]\neq0$, where $\hat{P}$ is the orthogonal
projection onto the truncated Fock space
($\hat{P}^{\dagger}=\hat{P}=\hat{P}^2$); $\hat{P}\hat{H}\hat{P}$ denotes
truncation. Bottom: the resulting QFI parabola, Eq.~\eqref{eq:main}, for
the harmonic oscillator at $\Omega/\omega=0.4$, $\eta=2$; only $\theta=0$
is attainable with a fixed apparatus.}
\label{fig:schematic}
\end{figure}

\smallskip\noindent\textit{Gauge orbit.} We use a one-dimensional model that contains the essential physics while keeping every operator exact,
\begin{equation}
\hat{H}_{\rm C}(g)=\frac{\hat{p}^2}{2}+V(\hat{x})+\omega\hat{a}^{\dagger}\hat{a}-g\,\hat{p}\hat{X}+\frac{g^2}{2}\hat{X}^2,
\qquad \hat{X}=\hat{a}+\hat{a}^{\dagger} ,
\label{eq:HC}
\end{equation}
i.e.\ the Coulomb-gauge, minimal-coupling form $(\hat{p}-g\hat{X})^2/2+V(\hat{x})+\omega\hat{a}^{\dagger}\hat{a}$. Here $\hat{p}$ is the momentum conjugate to $\hat{x}$, $V(\hat{x})$ the bare matter potential, and $\hat{a}^{\dagger}\hat{a}$ the photon number. Throughout, $\hbar=1$. The dipole gauge is reached with
\begin{equation}
\hat{H}_\theta(g)=\hat{U}_\theta^{\dagger}\hat{H}_{\rm C}\hat{U}_\theta,\qquad
\hat{U}_\theta=\exp\!\big(i\theta g\,\hat{x}\hat{X}\big),
\label{eq:orbit}
\end{equation}
where $\theta=1$ gives the dipole form used throughout the USC literature, and intermediate $\theta$ interpolates. Since $\hat{U}_\theta$ depends on $g$, Eq.~\eqref{eq:orbit} is not a change of basis. It maps the family $\hat{H}_{\rm C}(g)$ onto a different family. We stress what $\theta$ is and is not. For fixed $g$ the physical gauge transformation is the single point $\theta=1$. Intermediate $\theta$ corresponds to no electromagnetic gauge; we claim no continuous gauge freedom. Equation~\eqref{eq:orbit} is a one-parameter interpolation between two truncated models. It is the group action that would relate the frames exactly, were no truncation performed. The truncation, not any gauge symmetry, breaks their equivalence (Fig.~\ref{fig:schematic})~\cite{Rouse2021}. Interpolations of this kind have been used to optimize truncated models~\cite{Qing2025}. There the criterion is fidelity to exact static or dynamic observables. Here it is which bound an apparatus can reach. The two need not select the same $\theta$. We work at the single-mode level. Mode elimination and the fate of the $A^2$ coefficient~\cite{Rzazewski1975} raise separate issues. We use $\eta\equiv 2gx_{01}/\omega$, with $x_{01}=\langle 0|\hat{x}|1\rangle$ the dipole matrix element between the two lowest matter levels. The value $\eta\gtrsim0.1$ marks the onset of the USC regime. The effects below are largest in the deep-strong-coupling regime $g/\omega\gtrsim1$.

Two structural facts are used repeatedly. First, $[\hat{x}\hat{X},\hat{X}]=0$. The quadrature $\hat{X}$ and all its moments are therefore invariant along the orbit. Second, the spectrum of $\hat{H}_\theta$ is exactly $\theta$ independent. We use this as a numerical check.

We diagonalize Eq.~\eqref{eq:HC} in a basis of $N$ exact matter eigenstates $\ket{n}$ of the bare matter Hamiltonian $\hat{p}^2/2+V(\hat{x})$, tensor $n_{\rm ph}$ Fock states. The eigenvalues are $\varepsilon_n$. The projections $p_{nm}=i(\varepsilon_n-\varepsilon_m)x_{nm}$, with $x_{nm}=\langle n|\hat{x}|m\rangle$, are exact. Both cutoffs are converged to the digits shown. Four potentials are used, with anharmonicity $a_2=(\varepsilon_2-2\varepsilon_1+\varepsilon_0)/\omega$: harmonic ($a_2=0$), a flux-qubit-like cosine well ($a_2=-0.05$), a square well ($a_2=0.67$), and a double well ($a_2=2.42$). Throughout, $\Omega\equiv\varepsilon_1-\varepsilon_0$, normalized to $\Omega=\omega$ at zero detuning.

\begin{figure}[tb]
\includegraphics[width=\linewidth]{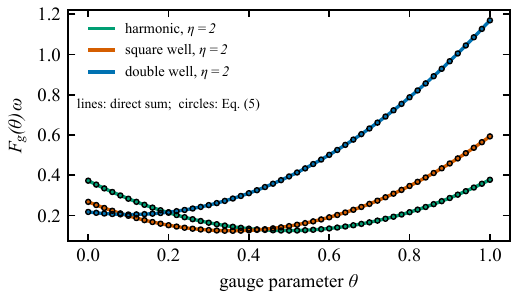}
\caption{(Color online) $F_g(\theta)$ along the orbit at resonance, $\eta=2$: lines, direct sum; circles, the closed form Eq.~\eqref{eq:main}. The harmonic case is exactly symmetric about $\theta_\ast=1/2$; the double well dips at $\theta_\ast\simeq0.1$ and rises by a factor $5.7$ to $\theta=1$.}
\label{fig:theta}
\end{figure}

\smallskip\noindent\textit{Exact gauge dependence of the QFI.} For a non-degenerate ground state $\ket{0}$ of $\hat{H}$, with eigenstates $\ket{n}$ and eigenvalues $E_n$,
\begin{equation}
F_g=4\sum_{n\ge1}\frac{|\langle n|\partial_g\hat{H}|0\rangle|^2}{(E_n-E_0)^2}.
\label{eq:qfi}
\end{equation}
Write $\ket{0_\theta}=\hat{U}_\theta^{\dagger}\ket{0_{\rm C}}$. Use $\hat{U}_\theta\,\partial_g \hat{U}_\theta^{\dagger}=-i\theta\hat{x}\hat{X}$ and $\hat{U}_\theta\hat{X}\hat{U}_\theta^{\dagger}=\hat{X}$. For every excited state one then finds
\begin{equation}
\langle n_\theta|\partial_g\hat{H}_\theta|0_\theta\rangle
=a_n+i\theta\,(E_n-E_0)\,x_n,
\label{eq:element}
\end{equation}
with $a_n=\langle n|\partial_g\hat{H}_{\rm C}|0\rangle$ and $x_n=\langle n|\hat{x}\hat{X}|0\rangle$ evaluated in the minimal-coupling frame. Substituting into Eq.~\eqref{eq:qfi} gives the central result:
\begin{equation}
F_g(\theta)=F_g(0)+4\theta^2{\rm Var}(\hat{x}\hat{X})-8\theta\,W,
\label{eq:main}
\end{equation}
\begin{equation}
W\equiv{\rm Im}\,\langle \Delta(\hat{x}\hat{X})\,\partial_g\psi\rangle ,
\qquad
\partial_g\psi=\sum_{n\ge1}\frac{\ket{n}\langle n|\partial_g\hat{H}_{\rm C}|0\rangle}{E_0-E_n}.
\label{eq:W}
\end{equation}
Here $\psi\equiv\ket{0_{\rm C}}$ is the ground state, $\Delta\hat{O}\equiv\hat{O}-\langle\hat{O}\rangle$, and hence ${\rm Var}(\hat{O})=\langle(\Delta\hat{O})^2\rangle$. Equation~\eqref{eq:main} is exact and non-perturbative. It involves only two properties of the ground state in the minimal-coupling frame: the variance of the gauge generator $\hat{x}\hat{X}$, and its covariance with the parameter derivative. We have verified it against Eq.~\eqref{eq:qfi} along the orbit for four potentials, several couplings, and all $\theta$. The relative residual is below $10^{-13}$ for the parameters of Fig.~\ref{fig:theta}, and below $10^{-6}$ elsewhere. The spectral cutoffs set this limit.

\emph{(i) The two frames are not equivalent.} At $\theta=1$,
$F_g^{\rm dip}/F_g^{\rm Coul}-1=\big[4\,{\rm Var}(\hat{x}\hat{X})-8W\big]/F_g^{\rm Coul}$. This is in general nonzero, and of either sign.

\emph{(ii) The gauge-invariant object requires a covariant derivative.} Since $\hat{U}_\theta^{\dagger}\partial_g\hat{U}_\theta=i\theta\hat{x}\hat{X}$ and $[\hat{x}\hat{X},\hat{U}_\theta]=0$, the combination $D_g=\partial_g+i\theta\hat{x}\hat{X}$ obeys $D_g\ket{0_\theta}=\hat{U}_\theta^{\dagger}\partial_g\ket{0_{\rm C}}$. The QFI computed with $D_g$ in place of $\partial_g$ is frame independent. It equals $F_g(0)$ in every gauge. The connection term $-i\theta\hat{x}\hat{X}$ is exactly what distinguishes a parameter-dependent change of family from a change of basis. Omitting it produces the difference~\eqref{eq:main}.

\emph{(iii) Closed form for the harmonic model.} For $V(\hat{x})=\Omega^2\hat{x}^2/2$ at $\omega=1$ the ground state is Gaussian, and everything closes. Write $\hat{\mathbf{r}}=(\hat{x},\hat{p},\hat{b},\hat{\Pi})^{\rm T}$, with $\hat{b}=\hat{X}/\sqrt{2}$ and $[\hat{x},\hat{p}]=[\hat{b},\hat{\Pi}]=i$. The ground-state covariance $\sigma=\tfrac12\langle\{\hat{r},\hat{r}^{\rm T}\}\rangle$ is then
\begin{equation}
\begin{aligned}
\sigma&=-\tfrac{1}{2}\,A\,(-A^2)^{-1/2}J,\\
A&=\begin{pmatrix}
0&1&-\sqrt{2}g&0\\
-\Omega^2&0&0&0\\
0&0&0&1\\
0&\sqrt{2}g&-(1+2g^2)&0
\end{pmatrix},
\end{aligned}
\label{eq:sigma}
\end{equation}
where $A$ is the drift matrix of the quadratic Hamiltonian, and $J$ the canonical symplectic form. By Cayley--Hamilton, $(-A^2)^{-1/2}=a_0+a_1(-A^2)$. The two coefficients follow from the normal-mode frequencies $\nu_\pm$ of $A$, $\nu_\pm^2=\tfrac12\big(c\mp\sqrt{c^2-4\Omega^2}\big)$ with $c=\Omega^2+1+2g^2$:
\begin{equation}
a_1=-\frac{1}{\Omega(\nu_++\nu_-)},\qquad
a_0=\frac{1}{\nu_+}-a_1\nu_+^2 .
\label{eq:ch}
\end{equation}
Equation~\eqref{eq:sigma} is exact; against exact diagonalization it agrees to $2\times10^{-16}$. The simultaneous reflection $(\hat{x},\hat{p},\hat{b},\hat{\Pi})\to-(\hat{x},\hat{p},\hat{b},\hat{\Pi})$ sets $\langle\hat{x}\rangle=\langle\hat{b}\rangle=0$. The reflection $S={\rm diag}(1,-1,-1,1)$ obeys $SAS=-A$. Hence $S\sigma=\sigma S$ and $\langle\hat{x}\hat{b}\rangle=0$. Wick's theorem then gives ${\rm Var}(\hat{x}\hat{X})=2\sigma_{xx}\sigma_{bb}$, in closed form for all $\Omega$ and $g$. At resonance $\Omega=1$ this collapses to ${\rm Var}(\hat{x}\hat{X})=1/(2+g^2)$ and $W=1/[2(2+g^2)]$, so that
\begin{equation}
F_g(\theta)=\frac{2(1+g^2)}{(2+g^2)^2}+\frac{4(\theta^2-\theta)}{2+g^2},
\label{eq:resF}
\end{equation}
an exact non-perturbative parabola. For the Gaussian ground state the two canonical frames agree identically: $F_g(1)=F_g(0)$. The symmetric intermediate gauge suppresses the QFI by $F_g(\theta_\ast)/F_g(0)=g^2/[2(1+g^2)]$, with $\theta_\ast=1/2$. Away from resonance the minimum moves. For the harmonic model it sits at
\begin{equation}
\theta_\ast=\frac{W}{{\rm Var}(\hat{x}\hat{X})}=\frac{\Omega}{\Omega+\omega},
\label{eq:thstar}
\end{equation}
independent of the coupling $g$; verified to $25$ significant digits at generic $(\Omega,g)$, and exact at resonance. The parabola is symmetric about it: $F_g(\theta)=F_g(2\theta_\ast-\theta)$. The minimum is thus the mean of the two canonical frames, weighted by the two frequencies. It approaches the minimal-coupling frame ($\theta=0$) for $\Omega\ll\omega$, and the dipole frame ($\theta=1$) for $\Omega\gg\omega$. Anharmonicity displaces it from this value (Fig.~\ref{fig:theta}). The minimum measures how far truncation displaces the bound, not which frame to work in. The physical value, argued below, is $F_g(0)$, not the extremum.

\begin{figure*}[t]
\includegraphics[width=0.82\textwidth]{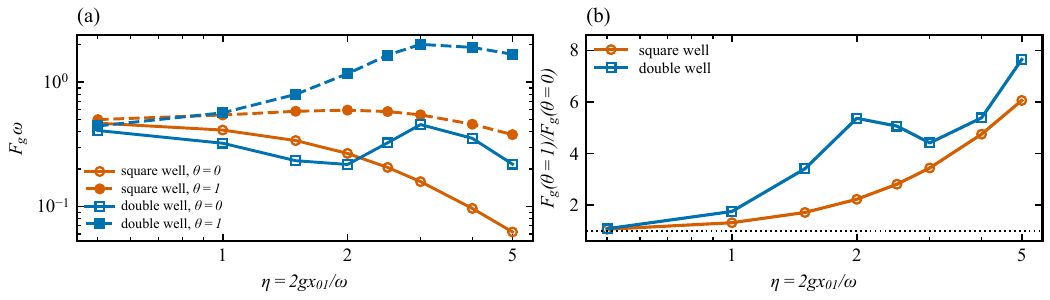}
\caption{(Color online) QFI for the coupling at resonance, $\Omega=\omega$. (a)~$F_g$ versus $\eta$ for the square well and the double well; solid: minimal-coupling frame ($\theta=0$), dashed: dipole frame ($\theta=1$). (b)~Ratio of the two.}\label{fig:1}
\end{figure*}

\smallskip\noindent\textit{Magnitude.} The disagreement is not a constant factor. For the square well [Fig.~\ref{fig:1}(a)], $F_g(\theta{=}0)$ falls by a factor $6.6$ between $\eta=1$ and $5$. The dipole-frame value falls by only $1.44$. The ratio grows from $1.32$ to $6.1$. For the double well the dipole-frame value even rises ($0.57\to1.68$), while the minimal-coupling value falls ($0.32\to0.22$). The ratio reaches $7.6$. The two gauges disagree not only on the magnitude, but on the scaling of the precision.

\begin{figure}[tb]
\includegraphics[width=\linewidth]{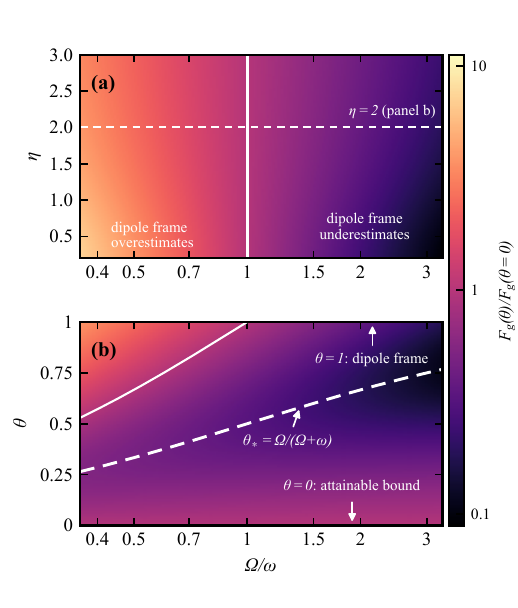}
\caption{(Color online) Harmonic oscillator, closed forms
Eqs.~\eqref{eq:sigma}--\eqref{eq:thstar}; no diagonalization. (a)~Ratio
$F_g(1)/F_g(0)$ in the $(\Omega/\omega,\eta)$ plane; the contour
marks exact equality, which for the harmonic oscillator lies at
$\Omega=\omega$. Bright: the dipole frame overestimates the attainable
precision; dark: it underestimates. (b)~$F_g(\theta)/F_g(0)$ in the
$(\Omega/\omega,\theta)$ plane at $\eta=2$ (dashed line above); solid
contour: $F_g(\theta)=F_g(0)$ at $\theta=2\theta_\ast$; dashed curve: the
valley $\theta_\ast=\Omega/(\Omega+\omega)$; lower edge $\theta=0$: the
attainable bound.}
\label{fig:detune}
\end{figure}

Detuning is as important as anharmonicity (Fig.~\ref{fig:detune}). The harmonic oscillator agrees exactly at resonance: $F_g(1)=F_g(0)$ [Eq.~\eqref{eq:resF}]. At $\Omega/\omega=0.4$, however, it gives $4.7$ at $\eta=1$ and $3.5$ at $\eta=2$. At $\Omega/\omega=3$ the sign reverses. The dipole frame then underestimates the precision, by up to a factor $6.4$. The largest discrepancy combines anharmonicity and detuning: $9.6$ for the double well at $\Omega/\omega=0.4$, $\eta=2$ ($g/\omega\simeq1$, within reach of current flux-qubit circuits~\cite{Yoshihara2017}). Across the scanned parameters the ratio spans $0.16$ to $9.6$. That is more than an order of magnitude, in either direction.

\smallskip\noindent\textit{A case study: critical metrology from virtual excitations.} The orbit formula bears directly on published USC metrology. Ref.~\cite{Hotter2025} works in the thermodynamic-limit Dicke model~\cite{Dicke1954} at resonance, and estimates the bare cavity frequency $\omega$ from ground-state virtual excitations. In the gauge-inconsistent form (dipole, no gauge-restoring $A^2$) the polariton frequencies $\omega_\pm$ split. The soft branch $\omega_-$ has dispersion
\begin{equation}
\omega_-^2=\tfrac12\Big[(\omega^2+1)-\sqrt{(\omega^2-1)^2+4\kappa^2\omega}\Big],
\label{eq:them}
\end{equation}
with $g_{\rm c}$ the critical coupling and $\kappa=g/g_{\rm c}$ the distance to it. The slope $\partial_\omega\omega_-=(2-\kappa)/[4\sqrt{1-\kappa}]$ then diverges as $\kappa\to1$, and the static ground-state QFI diverges as $(1-\kappa)^{-2}$. This is an apparent diverging metrological gain, funded by virtual excitations.

The gauge-consistent (Hopfield~\cite{Hopfield1958}) completion of the same model is not Eq.~\eqref{eq:them}. It is the minimal-coupling Hamiltonian of Eq.~\eqref{eq:HC} itself, with the bare matter frequency $\Omega$ and the Coulomb coupling $g_{\rm C}$ held fixed as $\omega$ varies. Its polaritons then obey the Rz\c{a}\.{z}ewski constraint~\cite{Rzazewski1975} as a protected product, $\omega_+^2\omega_-^2=\omega^2\Omega^2$, together with the sum $S(\omega)=\omega^2+\Omega^2+2\omega g_{\rm C}^2$. Matching the full polariton spectrum of Ref.~\cite{Hotter2025} at $\omega=1$ (sum $2$, product $1-\kappa^2$) fixes $\kappa=\sqrt{2}\,g_{\rm C}$ and $\Omega^2=1-\kappa^2$. The soft branch becomes
\begin{equation}
\omega_-^2=\tfrac12\Big[S(\omega)-\sqrt{S(\omega)^2-4\omega^2\Omega^2}\Big].
\label{eq:cons}
\end{equation}
The two spectra coincide at $\omega=1$, but only there. The truncated product $\omega_+^2\omega_-^2=\omega(\omega-\kappa^2)$ has lost the protection. Every derivative of the ground state feels the difference. The consistent slope $\partial_\omega\omega_-=(2-\kappa)\sqrt{1-\kappa}/4$ vanishes as $\kappa\to1$, instead of diverging. It does so for any definition of the family. What defines the family is which quantity the device holds fixed as $\omega$ varies. Take any family that keeps the bare matter frequency $\Omega$ fixed and matches the operating point of Ref.~\cite{Hotter2025} ($S(1)=2$, $\Omega^2=1-\kappa^2$). Differentiating the soft branch of Eq.~\eqref{eq:cons} at $\omega=1$, where $\omega_-=\sqrt{1-\kappa}$ and $\sqrt{S^2-4\omega^2\Omega^2}=2\kappa$, gives
\begin{equation}
\partial_\omega\omega_-\big|_{1}=\frac{\sqrt{1-\kappa}}{4\kappa}\big[2(1+\kappa)-S'(1)\big],
\label{eq:app}
\end{equation}
with $S'(1)=2+\kappa^2+2\,\partial_\omega g_{\rm C}^2|_{1}$. The prefactor $\sqrt{1-\kappa}$ is universal. Holding the Coulomb coupling fixed ($\partial_\omega g_{\rm C}^2=0$) recovers the slope above. Holding the dipole moment fixed ($g_{\rm C}\propto\omega^{-1/2}$, so $\partial_\omega g_{\rm C}^2=-\kappa^2/2$) gives $\sqrt{1-\kappa}/2$. Holding the dipole coupling fixed ($g_{\rm C}\propto\omega^{-1}$, so $\partial_\omega g_{\rm C}^2=-\kappa^2$) gives $(2+\kappa)\sqrt{1-\kappa}/4$. All three vanish as $\kappa\to1$. The truncated slope $(2-\kappa)/[4\sqrt{1-\kappa}]$ of Eq.~\eqref{eq:them} diverges. The correction is thus independent of what the device holds fixed. The dynamic-protocol gain of Ref.~\cite{Hotter2025} is reduced by the factor
\begin{equation}
R(\kappa)=(1-\kappa)^2,
\label{eq:R}
\end{equation}
and the static QFI no longer diverges but saturates (Table~\ref{tab:2}).

\begin{table}[b]
\caption{Estimating $\omega$ at the operating point of Ref.~\cite{Hotter2025}: truncated versus gauge-consistent family, with identical spectra at $\omega=1$. $R(\kappa)$: dynamic-protocol correction, Eq.~\eqref{eq:R}; $I_\omega$: static ground-state QFI from exact diagonalization of each quadratic family.}
\label{tab:2}
\begin{ruledtabular}
\begin{tabular}{lcccc}
$\kappa$ & $\partial_\omega\omega_-|_{\rm tr}$ & $\partial_\omega\omega_-|_{\rm cons}$ & $I_\omega^{\rm tr}$ & $I_\omega^{\rm cons}$ \\
\colrule
0.2  & 0.503 & 0.402 & 0.0028 & 0.0025 \\
0.5  & 0.530 & 0.265 & 0.036  & 0.017 \\
0.9  & 0.870 & 0.087 & 2.60   & 0.063 \\
0.99 & 2.53  & 0.025 & 305    & 0.072 \\
\end{tabular}
\end{ruledtabular}
\end{table}

In the consistent completion $\kappa\to1$ forces $\Omega\to0$. The ``critical point'' sits exactly on the Rz\c{a}\.{z}ewski boundary. It can be reached only by tuning the device away from it, and then the matter mode goes soft. The soft polariton then decouples from the parameter, and its slope vanishes. Absorbing the diamagnetic term into a redefinition of $g_{\rm c}$ does not evade this. Rescaling $\kappa$ is exactly what makes the two families agree at $\omega=1$ (Table~\ref{tab:2}). It cannot also erase the difference in their $\omega$ dependence, which is what $\partial_\omega\omega_-$ probes. The claimed critical enhancement of the QFI from virtual excitations is thus a gauge artifact of the truncated family. Equation~\eqref{eq:main} quantifies the same mechanism for the estimation of $g$ in finite systems. Here it acts on $\omega$, amplified by the proximity to an unphysical boundary.

\smallskip\noindent\textit{Which value is physical?} The QFI is the supremum over all positive-operator-valued measurements (POVMs). A POVM $\{\hat{E}_m\}$ fixed in the frame $\theta$ becomes $\{\hat{U}_\theta(g)\hat{E}_m\hat{U}_\theta^{\dagger}(g)\}$ in the minimal-coupling frame. That transformed set depends on the unknown $g$. The realizable measurements are therefore exactly those $g$-independent in that frame. Their supremum is $F_g(0)$. An apparatus may be designed with knowledge of the operating point. Once built, it is a fixed set of operators. The dipole-frame optimization instead lets $\hat{E}_m$ track $g$. No apparatus does that.

One premise deserves to be explicit. Fixing the apparatus in the minimal-coupling frame is not arbitrary. The question of how well $g$ can be estimated is frame independent. On the orbit only $F_g(0)$ is frame independent, as the covariant-derivative QFI of remark \emph{(ii)}. Every other point answers a question posed of a truncated model.

Two remarks. First, $\hat{X}$ is gauge invariant, so homodyne schemes are bounded by $F_g(0)$. The dipole-frame excess requires measurements not commuting with the generator $\hat{x}\hat{X}$. Second, the dipole gauge is standard in the USC literature because truncated models are usually found to be most accurate there spectrally~\cite{DeBernardis2018,DiStefano2019,StokesNazir2024}. Spectral convergence concerns eigenvalues, not the matrix elements $\langle n|\partial_g\hat{H}|0\rangle$ from which the bound is built. The price of that convenience is attainability.

\smallskip\noindent\textit{Discussion.} The result adds a second gauge ambiguity to the resource accounting of USC metrology. The mean ground-state photon number~\cite{DuttaKollar} and the figure of merit, Eq.~\eqref{eq:main}, both vary quadratically in the gauge parameter. Virtual excitations are claimed as a metrological resource in their own right. In Ref.~\cite{Hotter2025} the QFI scales with the square of the virtual-excitation number, $I_\omega\propto\langle\hat{n}_c\rangle^2$, termed Heisenberg scaling. This claim is frame dependent twice over. Its divergence does not survive gauge consistency (Table~\ref{tab:2}).

The effect is largest in the deep-strong-coupling regime, and it is decidable in a fixed device. The right side of Eq.~\eqref{eq:main} consists of ground-state moments: $\langle\hat{x}^2\rangle$, $\langle\hat{b}^2\rangle$, and $W$ from the flux, photon-number, and $g$ dependence of the response. Two truncations that fit one measured spectrum therefore predict a definite splitting of their bounds.

Equation~\eqref{eq:main} concerns a class, not one model. Any bound evaluated away from the minimal-coupling frame misestimates the attainable sensitivity by $4\theta^2{\rm Var}(\hat{x}\hat{X})-8\theta W$, of either sign. We restrict to a single mode, the ground state, and estimation of the coupling. Multimode and open-system extensions are left open.

\begin{acknowledgments}
\textbf{Acknowledgments.} Supported by NSFC (Grant No.~12205092) and the Hunan Natural Science Foundation and Education Department (Grant Nos.~2026JJ50100 and 24B0670).
\end{acknowledgments}

\vspace{1.5ex}
\noindent\textbf{Data availability.}---The data that support the findings of this Letter are not publicly available. The data are available from the authors upon reasonable request.

\end{document}